\documentclass{easdlw}
\usepackage{graphicx}\hoffset=3cm
\def\figcoD{%
\centerline{{\bf(a)}\hskip-15pt
\scalebox{0.34}{\includegraphics{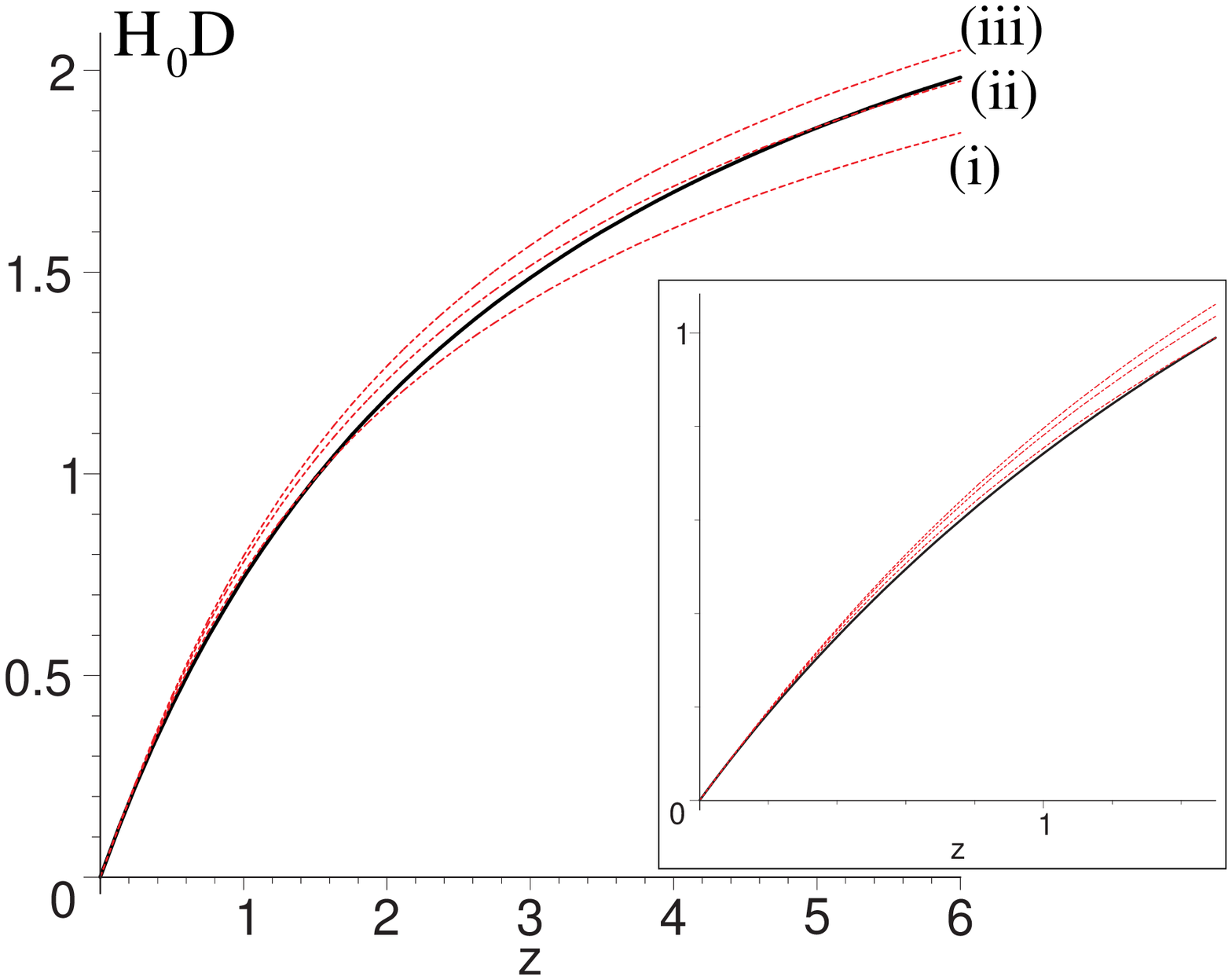}}
\quad{\bf(b)}\hskip-15pt\scalebox{0.34}{\includegraphics{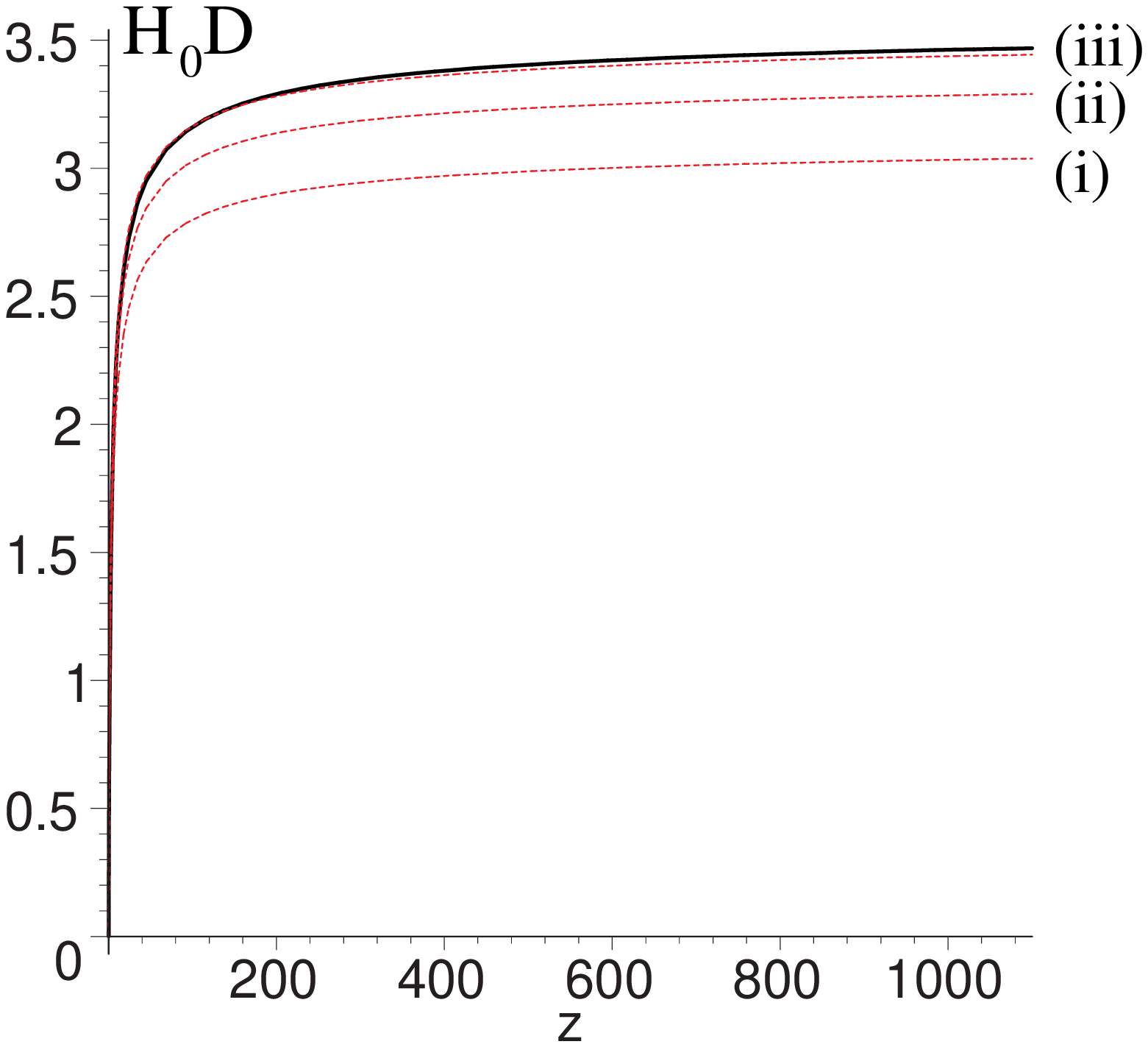}}}}
\def\PRL#1{Phys.\ Rev.\ Lett.\ {\bf#1}} \def\PR#1{Phys.\ Rev.\ {\bf#1}}
\def\ApJ#1{ApJ {\bf#1}}
\def\CQG#1{Class.\ Quantum Grav.\ {\bf#1}}
\def\GRG#1{Gen.\ Relativ.\ Grav.\ {\bf#1}}
\def\beq{\begin{equation}} \def\eeq{\end{equation}}
\def\bea{\begin{eqnarray}} \def\eea{\end{eqnarray}}
\def\Z#1{_{\lower2pt\hbox{$\scriptstyle#1$}}} \def\w#1{\,\hbox{#1}}
\font\sevenrm=cmr7 \def\ns#1{_{\hbox{\sevenrm #1}}} \def\dOM{\dd\Omega^2}
 \def\ave#1{\langle{#1}\rangle}
\def\lsim{\mathop{\hbox{${\lower3.8pt\hbox{$<$}}\atop{\raise0.2pt\hbox{$\sim$}}
$}}} \def\kmsMpc{\w{km}\;\w{sec}^{-1}\w{Mpc}^{-1}} 
\def\dd{{\rm d}} \def\ds{\dd s} \def\etal{{\em et al}.}
\def\al{\alpha}\def\be{\beta}\def\ga{\gamma}\def\de{\delta}
\def\th{\theta}\def\rh{\rho}\def\si{\sigma}
\def\ta{\tau} \def\ac{a}
\def\frn#1#2{{\textstyle{#1\over#2}}} \def\Deriv#1#2#3{{#1#3\over#1#2}}
\def\Der#1#2{{#1\hphantom{#2}\over#1#2}} \def\pt{\partial} \def\ab{{\bar a}}
\def\tw{\ta}\def\gb{\bar\ga}
\def\av{{a\ns{v}\hskip-2pt}} \def\aw{{a\ns{w}\hskip-2.4pt}}\def\Vav{{\cal V}}
\def\DD{{\cal D}}\def\gd{{{}^3\!g}}\def\half{\frn12}\def\Rav{\ave{\cal R}}
\def\QQ{{\cal Q}}\def\dsp{\displaystyle} \def\rw{r\ns w}
\def\mean#1{{\vphantom{\tilde#1}\bar#1}} 
\def\bH{\mean H}\def\Hb{\bH\Z{\!0}}\def\bq{\mean q}
\def\gb{\mean\ga}
\def\rhb{\mean\rh}\def\OM{\mean\Omega}\def\etb{\mean\eta}
\def\fw{{f\ns w}}\def\fv{{f\ns v}} \def\goesas{\mathop{\sim}\limits}
\def\fvn{f\ns{v0}} \def\fvf{\left(1-\fv\right)} \def\Hh{H}
\def\OMM{\OM\Z M}\def\OmMn{\Omega\Z{M0}} \def\ts{t} \def\Hm{H\Z0}
\def\Fi{\hbox{\footnotesize\it fi}}\def\etw{\eta\ns w} 
\def\fvi{{f\ns{vi}}} \def\fwi{{f\ns{wi}}}
\def\LCDM{$\Lambda$CDM} \def\OmLn{\Omega\Z{\Lambda0}}
\def\Dtc{\mathop{\hbox{$\Der\dd\tw$}}}
\begin{document}
\runningtitle{D.L.~Wiltshire: Gravitational Energy as Dark Energy}
\setcounter{page}{91}
\title{Gravitational energy as dark energy: Towards concordance cosmology
without Lambda}
\author{David L. Wiltshire}
\address{Department of Physics and Astronomy, University of Canterbury,
Private Bag 4800, Christ\-church 8140, New Zealand;
e-mail: {\tt David.Wiltshire@canterbury.ac.nz}}
\begin{abstract}
I briefly outline a new physical interpretation to the average cosmological
parameters for an inhomogeneous universe with backreaction. The variance
in local geometry and gravitational energy between ideal isotropic observers
in bound structures and isotropic observers at the volume average location in
voids plays a crucial role. Fits of a model universe to observational data
suggest the possibility of a new concordance cosmology, in which dark
energy is revealed as a mis-identification of gravitational energy gradients
that become important when voids grow at late epochs.
\end{abstract}
\maketitle
\section{Introduction}
The last decade has seen a shift in our understanding of the
expansion history of the universe, on account of precision measurements in
observational cosmology. The current prevailing view, based on model
universes which assume an exactly homogeneous and isotropic background
geometry for the universe, is that the universe is undergoing a period
of accelerating expansion, which began at relatively low redshifts. The
cause of this acceleration -- which in the standard framework would be
due to some smooth fluidlike dark energy component that violates the
strong energy condition -- is widely viewed as one of the biggest challenges
both for cosmology and fundamental physics. The simplest model for
dark energy -- a cosmological constant, $\Lambda$ -- is consistent with
many key observations, including in particular type Ia supernovae (SneIa)
luminosity distances,
the spectrum of cosmic microwave background (CMB) anisotropies, and
the echo of the baryon acoustic oscillation (BAO) scale in the primordial
plasma as reflected statistically in galaxy clustering.

Although the standard Lambda Cold Dark Matter (\LCDM) model provides
a good fit to many tests, there are tensions between some tests, and also
a number of puzzles and anomalies. Furthermore, theoretically
the existence of a cosmological constant begs the cosmic coincidence
question: why does the cosmological constant have a precise very tiny
value such that the universe only began accelerating at recent epochs,
making the matter density parameter, $\OmMn$, and cosmological constant
density parameter, $\OmLn$, of similar order today?

At the same time as the majority of cosmologists have been pursuing
ideas related to a fluidlike dark energy, or modifications of general
relativity, while retaining homogeneous isotropic backgrounds,
a small but growing number of cosmologists have questioned whether the
the expansion history of the universe may be understood in terms of the
growing inhomogeneous structure in recent epochs. (See, e.g., C\'el\'erier
(2007) for a review.) After all, at the present epoch the
universe is only statistically homogeneous once one samples on scales
of 150--300 Mpc. Below such scales it displays a web--like structure,
dominated in volume by voids. Some 40\%--50\% of the volume
of the present epoch universe is in voids of 30$h^{-1}$ Mpc (Hoyle \&
Vogeley 2002, 2004), where $h$ is the dimensionless parameter related to
the Hubble constant by $H\Z0=100h\kmsMpc$. Once one also accounts for
numerous minivoids, and perhaps also a few larger voids, then it appears that
the present epoch universe is void-dominated. Clusters of galaxies are spread
in sheets that surround these voids, and thin filaments that thread them.

One particular consequence of a matter distribution that is only
statistically homogeneous, rather than exactly homogeneous, is that when
the Einstein equations are averaged they do not evolve as a smooth
Friedmann--Lema\^{\i}tre--Robertson--Walker (FLRW) geometry. Instead
the Friedmann equations is supplemented by additional backreaction
terms (Buchert, 2000). Whether or not one can fully explain the expansion
history of the universe as a consequence of the growth of inhomogeneities
and backreaction, without a fluidlike dark energy, is the subject of
ongoing debate (Buchert, 2008). Over the past two years I have developed
a new physical interpretation of solutions to the Buchert equations
(Wiltshire 2007a, 2007b, 2008), with the conclusion that
a new concordance cosmology without exotic dark energy based on
a realistic average of the observed structures is a likely possibility.
In this paper I will briefly outline the key physical ingredients
of the new interpretation.

\section{Geometrical Averaging and Geometrical Variance}

The Buchert equations for irrotational dust (Buchert, 2000) involve spatial
averages on spacelike hypersurfaces. The equations take the form
\bea
&&\dsp{3\dot\ab^2\over\ab^2}=8\pi G\ave\rh-\half \Rav-\half\QQ,
\label{buche1}\\
&&{\ddot\ab\over\ab}=-4\pi G\ave\rh+\QQ,\label{buche2}\\
&&\pt_t\ave\rh+3{\dot\ab\over\ab}\ave\rh=0,\\
&&\pt_t\left(\ab^6\QQ\right)+\ab^4\pt_t\left(\ab^2\Rav\right)=0,
\label{buche3}\eea
where an overdot denotes a time derivative for observers comoving with the
dust of density $\rh$, $\ab(t)\equiv\left[\Vav(t)/\Vav(t\Z0)\right]^{1/3}$
with $\Vav(t)\equiv\int_\DD\dd^3x\sqrt{\det\gd}$, angle brackets denote the
spatial volume average of a quantity, and
$\QQ=\frn23\left(\langle\th^2\rangle-\langle\th\rangle^2\right)-
2\langle\si^2\rangle$,
is the kinematic backreaction, $\si^2=\frn12\si_{\al\be}\si^{\al\be}$ being
the scalar shear.

Since equations (\ref{buche1})--(\ref{buche3}) involve spatial averages,
their physical interpretation is not obvious. In particular $\ab(t)$ is not
the scale factor of a local metric, and the average spatial curvature, $\Rav$,
refers to a whole domain, $\DD$, on a spatial slice, rather an some more
local regional measurement. It is important to recall that in general
relativity we measure invariants of the {\em local} metric, not spatially
averaged quantities. If we are dealing with a genuinely inhomogeneous geometry,
with density contrasts $\de\rho/\rho\goesas-1$ on scales of 30$h^{-1}$ Mpc,
which is what is observed (Hoyle and Vogeley 2002, 2004) then we can
expect $\de{\cal R}/{\cal R}\goesas-1$ on similar scales.

Given such strong gradients in spatial curvature below the scale of
homogeneity, it is clear that not every observer is the same average
observer. Although average cosmic evolution may be governed by a set
of equations such as (\ref{buche1})--(\ref{buche3}), to physically
interpret their solutions we must consider where
the observers are within the inhomogeneous structure, and the physical
relationship of their local geometrical invariants to volume--average
ones. In other words, geometric variance can be just as important as
geometric averaging when it comes to the physical interpretation of the
expansion history of the universe. Any interpretation of averaged
inhomogeneous cosmologies which does not directly address this issue is
open to obvious potential criticisms (Ishibashi and Wald, 2006).

The physical interpretation of the Buchert equations I have developed is
based on the fact that structure formation provides a natural division
of scales in the observed universe. As observers in galaxies, we and the
objects we observe in other galaxies are necessarily in bound structures,
which formed from density perturbations that were greater than critical
density. If we consider the evidence of the large scale structure
surveys on the other hand, then the average location by volume in the
present epoch universe is in a void. If the presently observable universe
is underdense, a possibility that can arise by cosmic variance from an
initially near scale-free spectrum of density perturbations, then the voids
would have negative spatial curvature. There can therefore be systematic
differences of spatial curvature between the average mass environment, in
bound structures, and the volume-average environment, in voids.

\section{Gravitational Energy and Inhomogeneous Structure}

The definition of gravitational energy and conservation laws in general
relativity is extremely difficult, on account of the dynamical nature of
spacetime geometry and the equivalence principle. By the
strong equivalence principle, we can always get rid of gravity near a
point. However, those forms of energy which correspond to the kinetic energy
of expansion and to spatial curvature, which appear in the Einstein tensor
rather than the energy--momentum tensor, will generally have gradients
in an inhomogeneous universe. These regional quasilocal variations will affect
the relative calibration of clocks and rods at widely separated events.

In general, the question of how to synchronize clocks in the absence
of the exact symmetry described by a timelike Killing vector in general
relativity does not have a solution, and the definition of quasilocal
gravitational energy depends on choices of the splitting of spacetime
into spatial hypersurfaces, the threading of those hypersurfaces by observers,
and the associated choice of surfaces of integration. Such choices are
in general non-covariant and non-unique. One is essentially reduced to
asking which choices of frame have the greatest physical utility.

Since the ambiguities have their origin in the equivalence principle,
my view is that the equivalence principle should be properly
formulated and respected in the relative calibration of average frames in
cosmology. I have therefore extended the strong equivalence principle as
a cosmological equivalence principle (Wiltshire, 2008) to apply to average
spatially flat regions -- {\em cosmological inertial frames} -- undergoing a
regionally homogeneous isotropic volume expansion with deceleration over
arbitrarily long time intervals. By
thought experiments one can construct a Minkowski space analogue
for such frames, the semi-tethered lattice, by collectively applying brakes
in a synchronized fashion to freely unwinding tethers. In
special relativity, for two such lattices decelerating at
different rates, the observers in the lattice that decelerate more will age
less. By the cosmological equivalence principle, the same is true for
observers in expanding regions of different average density. Those
in the denser region decelerate more and age less. Since a relative
clock rate implies a gradient in gravitational energy, and a gradient in
average density a gradient in Ricci scalar curvature, this conceptually
establishes the notion of a gravitational energy cost for a spatial
curvature gradient. A small relative deceleration of the background,
typically of order $10^{-10}$ms$^{-2}$, cumulatively
leads to significant clock rate variances over the age of the universe
(Wiltshire, 2008).

By patching together cosmological inertial frames one obtains the
{\em cosmological rest frame}, namely an average global frame in which the
mean CMB temperature remains isotropic, even though the value of the mean
CMB temperature and the angular scale of the CMB anisotropies will vary with
changes in relative gravitational energy and spatial curvature from region to
region. The requirement for patching such regions together is that the
regionally measured expansion, in terms of the variation of the regional
proper length, $\ell_r=\Vav^{1/3}$, with respect to proper time of
isotropic observers (those who see an isotropic mean CMB), remains uniform.
Although voids open up faster, so that their proper volume increases more
quickly, on account of gravitational energy gradients the local clocks will
also tick faster in a compensating manner. This provides an
implicit solution to the Sandage--de Vaucouleurs paradox that a
statistically quiet, broadly isotropic, Hubble flow is observed deep below
the scale of statistical homogeneity.

The condition of an underlying uniform ``bare'' Hubble flow means that the
canonical isotropic observers will not in general be comoving with
dust on fine-grained scales. The Buchert average is taken to apply on
large scales, and to describe collective degrees of freedom of cells
which are coarse-grained at least on the size of statistical homogeneity.
The average scalar curvature of such cells and the average time parameter
are not assumed to coincide with the quantities locally measured by
isotropic observers within the cells. Ideally a new approach to cosmological
averaging, based on a uniform Hubble flow foliation, might be developed.
For the time being, I use the Buchert average equations for describing
cosmic evolution, while using the uniform local Hubble flow condition to
relate the volume--average quantities to parameters measured by observers
in spatially flat expanding {\em wall} regions containing galaxies.

Details of the fitting of local observables to average quantities for
solutions to Buchert's equations are described in detail in Wiltshire
(2007a, 2008). The model universe which is considered there is a
first approximation to the observed structures: negatively curved voids, and
spatially flat expanding wall regions within which galaxy clusters are
located, are combined in a Buchert average
\beq\fv(t)+\fw(t)=1,\eeq
where $\fw(t)=\fwi\aw^3/\ab^3$ is the {\em wall volume fraction} and
$\fv(t)=\fvi\av^3/\ab^3$ is the {\em void volume fraction},
$\Vav=\Vav\ns i\ab^3$ being the present horizon volume, and $\fwi$, $\fvi$ and
$\Vav\ns i$ initial values at last scattering. In trying to fit a
FLRW solution to the universe we attempt to
match our local spatially flat wall geometry
\beq\ds^2\Z{\Fi}=-\dd\tw^2+\aw^2(\tw)\left[\dd\etw^2+
\etw^2\dOM\right]\,.
\label{wgeom}\eeq
to the whole universe, when in reality the rods and clocks of ideal isotropic
observers vary with gradients in spatial curvature and gravitational energy.
By conformally matching radial null geodesics with those of the Buchert
average solutions, (\ref{wgeom}) may be extended to cosmological scales as
the dressed geometry
\beq
\ds^2=-\dd\tw^2+\ac^2(\tw)\left[\dd\etb^2+\rw^2(\etb,\tw)\,\dOM\right]
\label{dgeom}\eeq
where $a=\gb^{-1}\ab$, $\gb=\Deriv\dd\tw\ts$ is the relative lapse
function between wall clocks and volume--average ones, $\dd\etb=\dd t/\ab=
\dd\tw/ \ac$, and $\rw=\gb\fvf^{1/3}\fwi^{-1/3}\etw(\etb,\tw)$, where
$\etw$ is given by integrating $\dd\etw=\fwi^{1/3}\dd\etb/[\gb\fvf^{1/3}]$
along null geodesics.

In addition to the bare cosmological parameters which describe the Buchert
equations, one obtains dressed parameters relative to the geometry
(\ref{dgeom}). For example, the dressed matter density parameter is
$\Omega\Z M=\gb^3\OMM$, where $\OMM=8\pi G\rhb\Z{M0}\ab\Z0^3/(3\bH^2\ab^3)$
is the bare matter density parameter. The dressed parameters take numerical
values close to the ones inferred in standard FLRW models. Since the
relative lapse function is important in relating the bare and dressed
geometries, the interpretation is different to that of Buchert and Carfora
(2003), who considered dressing by volume factors relating to varying spatial
curvature.

\section{Apparent Acceleration and Apparent Hubble Flow Variance}

The gradient in gravitational energy and cumulative differences of clock
rates between wall observers and volume average ones has important
physical consequences. Using the exact solution to the Buchert equations
obtained in Wiltshire (2007b), one finds that a volume average observer
would infer an effective deceleration parameter $\bq=-\ddot\ab/(\bH^2\ab)=
2\fvf^2/(2+\fv)^2$, which is always positive since there is no global
acceleration. However, a wall observer infers a dressed deceleration
parameter
\beq
q=-{1\over H^2 a}{\dd^2 a\over\dd\tw^2}=
{-\fvf(8\fv^3+39\fv^2-12\fv-8)\over\left(4+\fv+4\fv^2\right)^2}\,,
\label{qtrack}\eeq
where the dressed Hubble parameter is given by
\beq\Hh=\ac^{-1}\Dtc\ac=\gb\bH-\dot\gb=\gb\bH-\gb^{-1}\Dtc\gb\,.
\label{42}\eeq
At early times when $\fv\to0$ the dressed
and bare deceleration parameter both assume the Einstein--de Sitter value
$q\simeq\bq\simeq\half$. However, unlike the bare parameter which
monotonically decreases to zero, the dressed parameter becomes negative
when $\fv\simeq0.59$ and $\bq\to0^-$ at late times. For the best-fit
parameters (Leith, Ng and Wiltshire, 2008) the apparent acceleration begins
at a redshift $z\simeq0.9$.

Cosmic acceleration is thus revealed as an apparent effect which arises
due to the cumulative clock rate variance of wall observers relative to
volume--average observers. It becomes significant only when the voids
begin to dominate the universe by volume. Since the epoch of onset of
apparent acceleration is directly related to the void fraction, $\fv$, this
solves one cosmic coincidence problem.

In addition to apparent cosmic acceleration, a second important apparent
effect will arise if one considers scales below that of statistical
homogeneity. By any one set of clocks it will appear that voids expand
faster than wall regions. Thus a wall observer will see galaxies on the
far side of a dominant void of diameter $30h^{-1}$ Mpc to recede at a
value greater than the dressed global average $\Hm$, while galaxies within
an ideal wall will recede at a rate less than $\Hm$. Since the uniform
bare rate $\bH$ would also be the local value within an ideal wall, eq.\
(\ref{42}) gives a measure of the variance in the apparent Hubble flow.
The best fit parameters (Leith, Ng and Wiltshire, 2008) give a dressed Hubble
constant $\Hm=61.7^{+1.2}_{-1.1}\kmsMpc$, and a bare Hubble constant
$\Hb=48.2^{+2.0}_{-2.4}\kmsMpc$. The present epoch variance is 22\%, and
we can expect the Hubble constant to attain local maximum values of order
$75\kmsMpc$ when measured over local voids.

Since voids dominate the universe by volume at the present epoch, any
observer in a galaxy in a typical wall region will measure locally higher
values of the Hubble constant, with peak values of order $75\kmsMpc$ at the
$30h^{-1}$ Mpc scale of the dominant voids. Over larger distances, as the
line of sight intersects more walls as well as voids, a radially spherically
symmetric average will give an average Hubble constant whose value decreases
from the maximum at the $30h^{-1}$ Mpc scale to the dressed global average
value, as the scale of homogeneity is approached at roughly the BAO scale of
$110h^{-1}$Mpc. This predicted effect would account for the Hubble bubble
(Jha, Riess \& Kirshner, 2007) and more detailed studies of the scale
dependence of the local Hubble flow (Li and Schwarz, 2008).

In fact, the variance of the local Hubble flow below the scale of homogeneity
should correlate strongly to observed structures in a manner which has no
equivalent prediction in FLRW models. This would provide a definitive
test of the proposal. It would also suggest that one contributing factor
in the decades long debate about the value of the Hubble constant is
the scale of averaging.

\section{Present and Future Observational Tests}

In addition to tests below the scale of homogeneity, equivalents to
the all standard cosmological tests can be derived. Fits to CMB
anisotropy data require that standard numerical codes must first be
rewritten from first principles, as the assumption of a homogeneous
isotropic cosmology is built into existing codes in a fundamental
way. As a first test of the CMB data, the fit to the angular scale
of the sound horizon has been examined (Wiltshire 2007a; Leith, Ng and
Wiltshire, 2008). It is often said that the overall angular scale of CMB
anisotropy spectrum, and in particular of the first Doppler peak, is
a measure of the spatial curvature of the universe. However, this is
only true if one assumes that the spatial curvature of the universe is
the same everywhere and that the universe evolves by the Friedmann equation.
In the presence of strong inhomogeneities and spatial curvature gradients
the analysis must be redone.

In Wiltshire (2007a) the angular scale of the sound horizon was analysed,
accounting for the fact that a volume--average observer in a void sees a
cooler mean CMB temperature, and dressed matter density and Hubble parameters
were found. Interestingly, since the calibration of the baryon to photon
ratio is affected, one finds that typically one can accommodate parameter
values which would agree with measurements of primordial lithium abundances
-- a problem for the \LCDM\ concordance cosmology --
while at the same time having more baryons relative to nonbaryonic dark
matter, as is required to fit the ratio of the heights of the first and
second Doppler peaks. This of course can only be confirmed once a full
numerical analysis of all the Doppler peaks is performed.

One can similarly determine parameter values which would give a dressed
comoving scale corresponding to the BAO scale as seen in galaxy clustering
statistics. Again, a detailed analysis would require the full use of the
new cosmological model in the data reduction. However, as a first approximation
one can assume the scale is the same as observed in the spatially flat \LCDM\
model. One finds that parameter values for which the angular scale of the
sound horizon and the comoving BAO scale agree with each other are the same
parameter values that best fit the Riess \etal\ (2007) (Riess07) SNeIa gold
dataset (Leith, Ng \& Wiltshire, 2008). Interestingly, such concordance is
found for a value of the dressed Hubble constant which agrees with that of
Sandage \etal\ (2006).

\pagestyle{addnote}
In Wiltshire (2008) several potential cosmological tests are investigated. It
is expected that the next generation of dark energy experiments, designed
to measure cosmological parameters to high accuracy, should also be able to
distinguish the new cosmological model from standard FLRW models.
As one example, in Fig.~1 we plot the effective comoving distance to a
redshift $z$ for the best-fit nonlinear bubble model, as compared to three
spatially flat \LCDM\ models with different values of $\OmMn$. At low
redshifts it closely matches the LCDM model which best fits the Riess07
SneIa gold data set only, while at high redshifts it closely matches the
\LCDM\ model which best fits WMAP5 (Komatsu \etal, 2009) only.
\begin{figure}[htb]
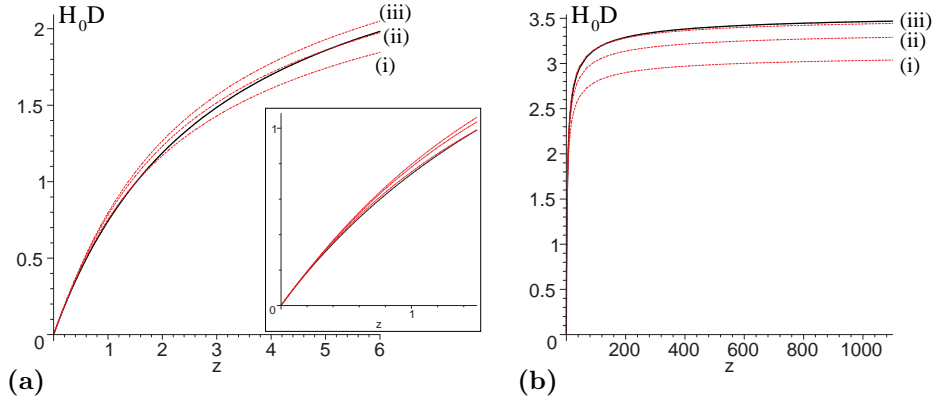

\vbox{\figcoD 
\caption{\label{fig_coD}$^*$
{\sl The effective comoving distance $\Hm D(z)$ is plotted for
the best--fit nonlinear bubble model, with $\fvn=0.762$, (solid line); and
for various spatially flat \LCDM\ models (dashed lines). The parameters for
the dashed lines are (i) $\OmMn=0.34$ (best--fit to SneIa only); (ii)
$\OmMn=0.279$ (joint best--fit to SneIa, BAO and WMAP5); (iii) $\OmMn=0.249$
best--fit to WMAP5 only. Panel {(a)} shows the redshift range $z<6$, with
an inset for $z<1.5$, the range tested by current SneIa data.
Panel {(b)} shows the range $z<1100$ up to the surface of last scattering,
tested by WMAP5.}}}
\end{figure}

\end{document}